\documentclass[conference]{IEEEtran}
\IEEEoverridecommandlockouts
\usepackage{cite}
\usepackage{amsmath,amssymb,amsfonts}
\usepackage{algorithmic}
\usepackage{graphicx}
\usepackage{textcomp}
\usepackage{xcolor}
\usepackage{multirow}
\usepackage{graphicx}
\usepackage{subcaption} 
\usepackage{adjustbox}   % Helps in centering subfigures
\usepackage[utf8]{inputenc}
\usepackage{newunicodechar}
\newunicodechar{✗}{\ding{55}} % or {\texttimes}, or just X
\usepackage{pifont} % for \ding{55}
\usepackage{verbatim}
\usepackage{fancyvrb}
\usepackage{url}
\usepackage{tikz}
\usetikzlibrary{arrows.meta, positioning, shapes, calc, fit}
\usepackage{pifont}% http://ctan.org/pkg/pifont
\newcommand{\cmark}{\ding{51}}  % ✓
\newcommand{\xmark}{\ding{55}}  % ✗
\usepackage[most]{tcolorbox}
\usepackage{float} % For floating box with !t
\usepackage{placeins}
\usepackage{tabularx}
\tcbset{
  ddossummary/.style={
    floatplacement=!ht,
    float,
    colback=red!5!white,
    colframe=black,
    boxrule=0.5pt,
    arc=4pt,
    left=6pt,
    right=6pt,
    top=6pt,
    bottom=6pt,
    enhanced,
    sharp corners=south,
    before skip=10pt,
    after skip=10pt,
    title={\textbf{DDoS Detection Model Output Summary}},
  }
}

\def\BibTeX{{\rm B\kern-.05em{\sc i\kern-.025em b}\kern-.08em
    T\kern-.1667em\lower.7ex\hbox{E}\kern-.125emX}}
\begin{document}

\title{Agentic AI for 6G: A New Paradigm for Autonomous RAN Security Compliance}

\author{
\IEEEauthorblockN{
Sotiris Chatzimiltis\IEEEauthorrefmark{1}, \textit{Graduate Student Member, IEEE},
Mahdi Boloursaz Mashhadi\IEEEauthorrefmark{1}, \textit{Senior Member, IEEE}, \\
Mohammad Shojafar\IEEEauthorrefmark{1}, \textit{Senior Member, IEEE},
Merouane Debbah\IEEEauthorrefmark{2}, \textit{Fellow, IEEE}, 
and Rahim Tafazolli\IEEEauthorrefmark{1}, \textit{Fellow, IEEE}
}
\IEEEauthorblockA{
\IEEEauthorrefmark{1}5G/6GIC, Institute for Communication Systems (ICS), University of Surrey, Guildford, UK\\
\{sc02449, m.boloursazmashhadi, m.shojafar, r.tafazolli\}@surrey.ac.uk
}
\IEEEauthorblockA{
\IEEEauthorrefmark{2}6G Research Center, Khalifa University, Abu Dhabi, UAE\\
merouane.debbah@ku.ac.ae
}
}
\maketitle

\begin{abstract}
Agentic AI systems are emerging as powerful tools for automating complex, multi-step tasks across various industries. One such industry is telecommunications, where the growing complexity of next-generation radio access networks (RANs) opens up numerous opportunities for applying these systems. Securing the RAN is a key area, particularly through automating the security compliance process, as traditional methods often struggle to keep pace with evolving specifications and real-time changes. In this article, we propose a framework that leverages LLM-based AI agents integrated with a retrieval-augmented generation (RAG) pipeline to enable intelligent and autonomous enforcement of security compliance. An initial case study demonstrates how an agent can assess configuration files for compliance with O-RAN Alliance and 3GPP standards, generate explainable justifications, and propose automated remediation if needed. We also highlight key challenges such as model hallucinations and vendor inconsistencies, along with considerations like agent security, transparency, and system trust. Finally, we outline future directions, emphasizing the need for telecom-specific LLMs and standardized evaluation frameworks. 
\end{abstract}


\section{Introduction}\label{sec:introduction}
The telecommunications industry is rapidly adopting artificial intelligence (AI)–driven solutions to manage the growing complexity and scale of next-generation mobile networks. Traditional machine learning (ML) and deep learning (DL) techniques have demonstrated strong performance in tasks such as resource allocation, network optimization, and anomaly detection. However, next-generation networks are increasingly evolving toward an intent-driven paradigm, wherein operators specify high-level goals rather than explicit configuration parameters. In such environments, existing ML and DL methods often lack the contextual understanding needed to interpret operator intents. 
%mekrache2024intent
To address these emerging challenges, recent research has turned to large language models (LLMs) as a promising new paradigm, offering advanced reasoning, interpretability, and contextual awareness that can support intelligent, adaptive network management and automation. Yet, despite their versatility, LLMs trained on generic datasets often lack the domain-specific knowledge and technical precision required for telecommunications applications. To fully harness their potential, there is a growing need for telecommunications foundation models trained on industry-relevant datasets and aligned with standardization requirements~\cite{maatouk2024large,zhou2024large}.  %bariah2024large,

Among the emerging applications of LLMs in telecommunications is network security. Researchers are increasingly investigating how LLMs can enhance threat modeling, vulnerability assessment, and anomaly detection. In the context of 5G and 6G networks, efforts have integrated LLMs into the STRIDE threat modeling framework to automate threat classification and recommend appropriate mitigations~\cite{abdulghaffar2025llms}. In parallel, LLMs have been proposed as autonomous reasoning engines capable of analyzing security risks, correlating multi-layer telemetry, and dynamically adapting security policies in real time~\cite{nguyen2024large}. Recent studies have further applied LLM-based approaches within Open RAN (O-RAN) and AI-RAN environments, demonstrating their potential for explainable intrusion detection~\cite{chatzimiltis2025xaillm}.
%and secure slicing ,moore2025integrated
Despite the advanced reasoning capabilities inherent in LLMs, their standalone use faces fundamental limitations that restrict their effectiveness for deployment in real-world telecommunication environments. A primary constraint is the static nature of their pre-trained knowledge, which limits their ability to incorporate real-time network states, evolving standards, and domain-specific data. Furthermore, LLMs are fundamentally reactive, requiring explicit human prompting for every action, which limits their suitability for dynamic environments demanding autonomous, time-critical decisions. 

To overcome these constraints, the paradigm of agentic AI has emerged. In this paradigm, LLMs operate within autonomous agents capable of defining goals, invoking tools, leveraging real-time information, and executing actions with minimal human intervention. This shift from static, prompt-driven models to autonomous, context-aware agents represents a promising direction for intelligent decision-making within next-generation RANs.

\begin{table*}[!t]
\centering
\scriptsize
\setlength{\tabcolsep}{2pt}
\caption{Checklist Comparison of Agentic AI Research in 6G, AI-RAN, and O-RAN.}
\label{tab:agentic_checklist}

\begin{tabularx}{\textwidth}{lXccccc}
\hline
\textbf{Ref.} &
\textbf{Problem Domain \& Solution Approach} &
\textbf{Security Focus} &
\textbf{Standards Aligned} &
\textbf{Multi-Agent} &
\textbf{RAG} \\
\hline

\cite{dev2025adv} &
6G architectures and V2X optimization via multi-agent reasoning &
\xmark &
\xmark &
\cmark &
\cmark \\

\cite{sharma2025mobillm} &
O-RAN threat mitigation using multi-agent LLMs with FiGHT-guided closed-loop responses &
\cmark &
\xmark &
\cmark &
\cmark \\

\cite{nezami2025standardization} &
Standardization of GenAI-driven agentic architectures for RAN &
\xmark &
\cmark &
\cmark &
\cmark \\

\cite{elkael2025agentran} &
Intent-driven autonomous RAN control via multi-layer LLM-based agents &
\xmark &
\xmark &
\cmark &
\xmark \\

\cite{salama2025edge} &
Edge-based autonomous RAN optimisation using persona-driven agents with predictive LSTM tools &
\xmark &
\xmark &
\cmark &
\xmark \\

\cite{wu2025llm} &
Adaptive O-RAN slice resource optimization using an LLM-driven agentic xApp &
\xmark &
\xmark &
\xmark &
\xmark \\

\cite{chatzistefanidis2025mxaiagenticobservabilitycontrol} &
Intent-driven O-RAN observability and configuration control via multi-agent planning and telemetry retrieval &
\xmark &
\xmark &
\cmark &
\cmark \\

\textbf{Our Work} &
\textbf{AI-RAN Security compliance automation using standards-aware multi-agent LLM reasoning} &
\textbf{\cmark} &
\textbf{\cmark} &
\textbf{\cmark} &
\textbf{\cmark} \\
\hline
\end{tabularx}

\vspace{1mm}
\begin{minipage}{0.97\textwidth}
\scriptsize
\textbf{Legend:} \cmark = supported; \xmark = not supported.  
“Standards Aligned” denotes explicit reasoning grounded in 3GPP and O-RAN specifications.
\end{minipage}

\end{table*}

Consequently, recent works have begun to investigate agentic AI across the 6G landscape, from architectural design to security and resource optimization. At the architectural level, recent research positions agentic AI as a key building block for AI-native 6G networks, enabling intent-driven orchestration, multi-domain automation, and streamlined control/user plane interactions~\cite{dev2025adv}. The same paradigm is gaining traction in the security domain, where agentic frameworks are used for closed-loop threat mitigation within O-RAN environments~\cite{sharma2025mobillm}. Standardization efforts are converging toward this direction, with emerging work defining how agentic intelligence should be embedded into O-RAN and AI-RAN architectures~\cite{nezami2025standardization}. In parallel, architectural proposals such as AgentRAN illustrate how hierarchical teams of LLM-based agents can operationalize intent-driven, multi-timescale control across open 6G networks~\cite{elkael2025agentran}. Complementary system-level studies further demonstrate that agentic AI can support resource optimization~\cite{salama2025edge}, enhance RAN resilience~\cite{wu2025llm}, and enable distributed observability and coordinated control across heterogeneous RAN components~\cite{chatzistefanidis2025mxaiagenticobservabilitycontrol}.

%,oran2024platformsecurity
Despite this rapid progress, current efforts remain largely focused on optimization and threat mitigation, leaving broader aspects of security assurance comparatively unexplored. In line with ongoing standardization efforts across ITU, 3GPP, and the O-RAN Alliance~\cite{itu2025genaitelecom,oran2025genai}, our work positions agentic AI as a key enabler for trustworthy, explainable, and autonomous security assurance, an essential capability for customer-centric, AI-native 6G networks. As RAN architectures evolve, ensuring that network components and services conform to changing security requirements necessitates both specification-aware intelligence (i.e., documentation-based compliance) and runtime observability (i.e., dynamic, evidence-driven monitoring). Traditional compliance methods, which rely mainly on static documentation and offline analysis, are insufficient in environments where configurations evolve continuously through intelligent applications. Meeting this challenge calls for a unified compliance model that spans both \textbf{compliance by design} and \textbf{compliance by evidence}~\cite{salman2024compliance}. Such a model requires systems not only to demonstrate conformance to formal standards specifications but also to provide contextual, verifiable justifications for their behaviour at runtime.

\noindent The article aims to answer the following questions:
\begin{itemize}
    \item How agentic AI entities are placed across the main functional layers of AI-RAN, how they interact and exchange information, and which security-focused use cases they can enable.
    \item In what ways LLM-enabled agentic AI systems can interpret and operationalize telecom security standards (e.g., 3GPP, O-RAN) to support an end-to-end compliance framework that integrates specification-aware reasoning with dynamic, evidence-based monitoring.
    \item To what extent an AI-agent prototype can assess configuration artifacts, identify security non-compliance, generate explainable justifications, autonomously propose remediation, and what challenges and deployment considerations arise in real-world RAN environments.
\end{itemize}

\section{Agentic AI in Next-Gen RAN: Architecture and Security Use Cases}
\label{sec:agentic_ai_ran}

Next-generation RANs are increasingly adopting software-defined, virtualized, and disaggregated designs, most notably O-RAN and AI-RAN. These architectures improve modularity and programmability, but also expand the network threat surface. In such dynamic environments, network functions must be adaptive and respond to changes in topology, traffic patterns, slice configurations, and intelligent application behavior. Conventional AI/ML approaches are effective for tasks such as slicing, resource allocation, and anomaly detection, but are typically trained offline and require human oversight. Agentic AI, as a possible solution, addresses these limitations by introducing autonomous entities capable of executing closed-loop cycles of \textit{observation}, \textit{reasoning}, and \textit{action}. Embedded within the RAN architecture, these agents can interpret live network states, enforce policies, and coordinate responses to security or performance deviations. Through continuous sensing and adaptation, agentic AI paves the way towards self-optimizing, zero-touch network operation.  

\begin{figure}[!t]
    \centering
    \includegraphics[width=\columnwidth]{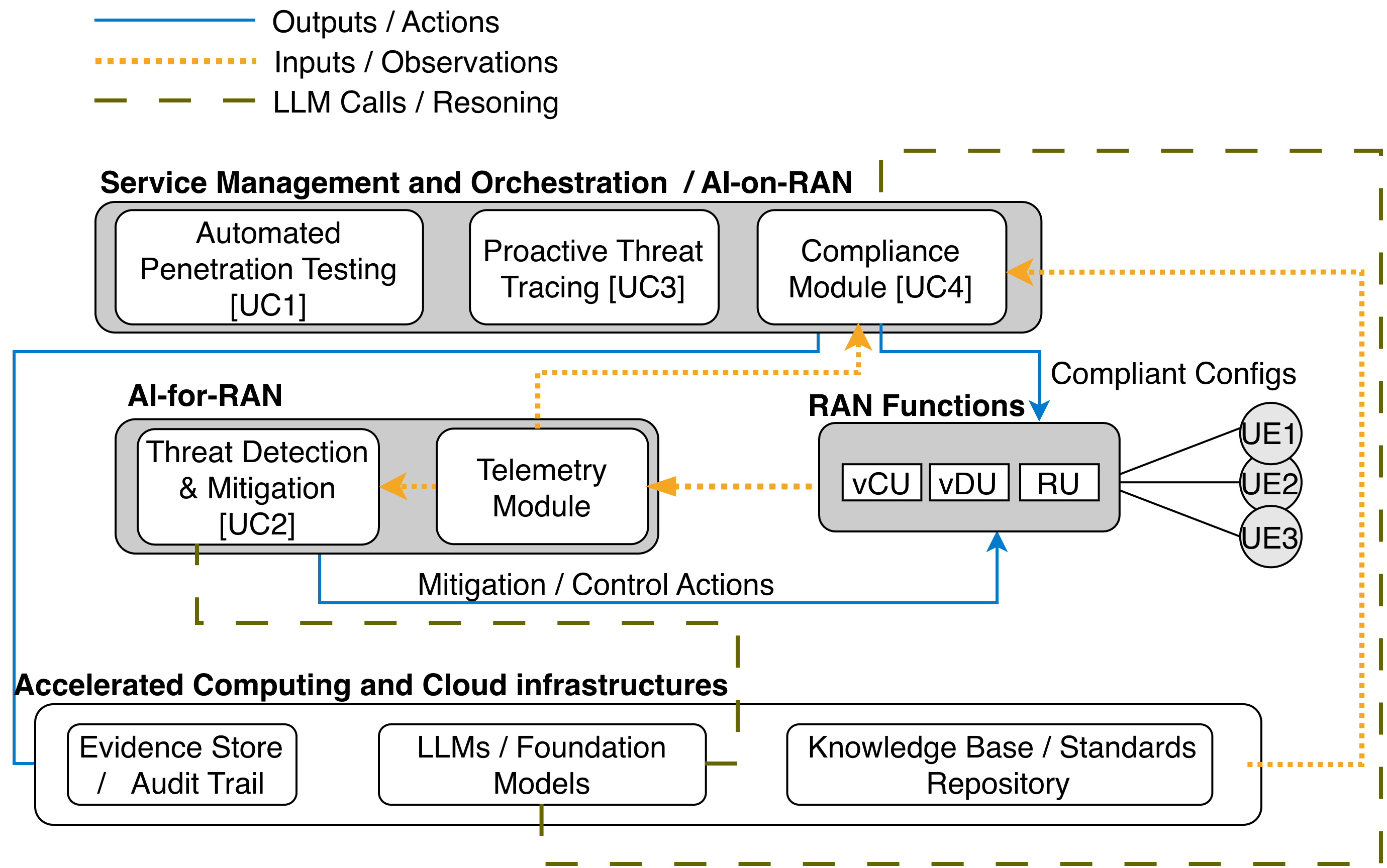}
    \caption{Placement of agentic AI entities and information flows across the AI-RAN architecture}
    \label{fig:openran_agenticai}
\end{figure}

Figure~\ref{fig:openran_agenticai} illustrates an example agentic AI-RAN architecture, depicting the placement and interaction of intelligent agents across the main functional layers. The upper layer (AI-on-RAN) hosts orchestration, management, and security agents, including penetration testing, threat tracing, and compliance modules. The middle layer (AI-for-RAN) integrates AI-driven computing functions, including telemetry processing and threat detection and mitigation, which interface with the virtualized RAN components (vCU, vDU, RU) to provide near real-time responses. At the foundation, the cloud and AI infrastructure layer provides LLMs, knowledge bases, and evidence stores that support reasoning, policy generation, and compliance verification.

The figure also highlights data and control flows among these components: orange dotted lines represent input data and telemetry consumed by agents, solid blue lines indicate outputs and control actions, and green dashed lines denote reasoning and verification interactions with the cloud-hosted foundation models. Together, these exchanges enable a continuous \textit{observe–reason–act} cycle, advancing toward autonomous RAN operation.

\subsection{Use Cases of Agentic AI for Securing Next-Gen RANs}
Within this architecture, agentic AI systems can play a critical role in both security operations (SecOps) and application security (AppSec). Operating continuously and with contextual awareness, these agents automate risk management processes spanning testing, monitoring, and response. The following use cases illustrate how agentic AI enhances the resilience and trustworthiness of next-generation RANs.

\begin{figure*}[!t]
    \centering
    \includegraphics[width=\textwidth]{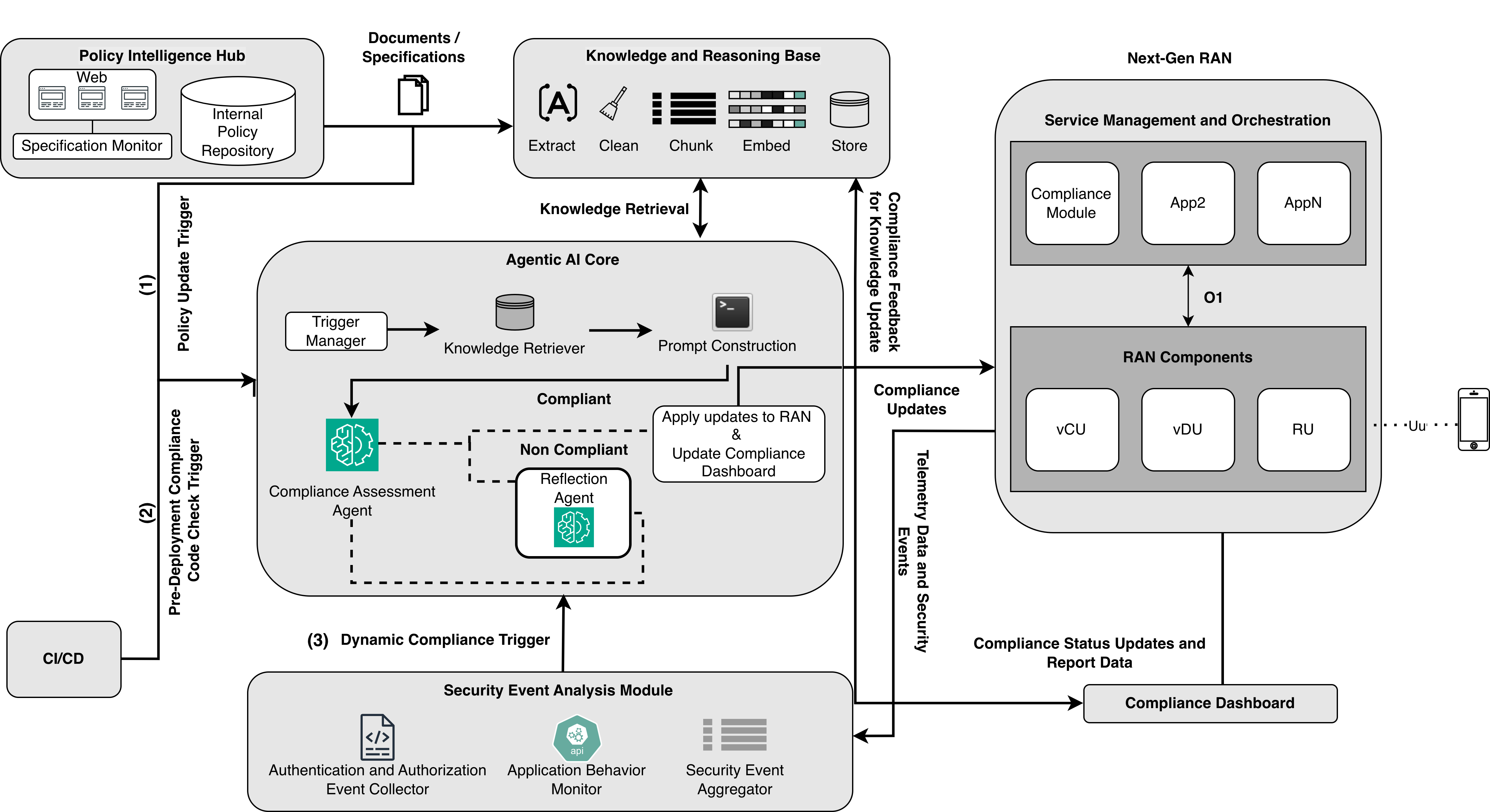}
    \caption{Proposed framework for intelligent security compliance in next-generation RANs.}
    \label{fig:proposed_framework}
\end{figure*}

\begin{itemize}
    \item \textbf{Intelligent Application Automated Penetration Testing:} Agentic AI can be deployed to autonomously conduct penetration testing on intelligent applications (e.g., dApps, xApps, and rApps) operating within next-generation RANs. These agents are capable of simulating adversarial behavior, dynamically adapting their testing strategies based on recent code changes or updated threat models, and continuously assessing applications for vulnerabilities. Unlike traditional penetration testing, which is periodic and manual, agentic AI enables continuous, context-aware testing integrated into CI/CD pipelines.
    
    In next-generation RANs where intelligent applications change dynamically, ensuring a consistent security posture is critical. For example, upon detecting the deployment of a new xApp, an AI agent could automatically initiate a series of security evaluations, testing the application’s API surface, data-handling routines, and integration points for exploitable weaknesses. The agent can then generate detailed security reports, recommend remediation actions, or even autonomously apply software patches. This continuous and automated approach ensures that applications are secured by design and maintain security throughout their lifecycle.

    \item \textbf{Real-Time Threat Detection and Mitigation Systems:} Agentic AI can enhance the capabilities of intrusion detection systems by continuously analyzing real-time telemetry across multiple layers of the RAN (i.e., physical, transport, and application). Rather than simply flagging anomalies, such as traffic surges indicative of DDoS attacks or rogue UE behavior, these agents could add contextual understanding to each event. Their key strength, though, lies in their ability to autonomously initiate mitigation workflows, such as rerouting traffic, isolating affected components, or enforcing access control restrictions. This reduces threat containment response time and minimizes human intervention, paving the way toward resilient, self-healing RAN environments. 
   % \textcolor{blue}{For the future challenges, we need to consider false positives, what happens if we fully automate the mitigation processes, can this lead to a catastrophe?}
    
    %same as adaptive threat hunting 
    \item \textbf{Proactive Threat Tracing:} Beyond real-time detection, agentic AI can support proactive threat tracing, also known as threat hunting, by continuously scanning the network for subtle indicators of compromise that may evade monitoring systems.  These agents analyze a wide range of telemetry, including logs, network activity, and endpoint behaviors, to identify indicators (e.g., IP addresses) and behavioral patterns (e.g., suspicious communication flows or unauthorized access attempts). By autonomously generating and executing queries across historical security data sources, agentic AI correlates diverse observations to uncover emerging threats before they escalate.

    \item \textbf{Security Compliance:} Agentic AI can serve as a continuous compliance enforcer within next-generation RANs, where the dynamic deployment of components and services complicates adherence to evolving security standards. These agents could interpret formal standards specifications, such as those from 3GPP, NIST, O-RAN, or AI-RAN and continuously validate the live network state against these requirements.

    Rather than relying on manual audits or static configuration checks, agentic AI can monitor both design-time intents and runtime behaviors, ensuring that deployed network functions comply with established requirements. For instance, when a new virtualized RAN component is provisioned, an AI agent could automatically validate that encryption protocols are correctly applied, and that access control rules are enforced. Importantly, the agent can detect misconfigurations or violations in real time and alert operators or autonomously initiate corrective actions. Finally, these systems could also contribute to operational transparency by generating explainable compliance evidence, such as annotated logs, compliance scores, or reasoning traces, which can support regulatory audits and post-incident forensics.
\end{itemize}

\section{CASE STUDY: AGENTIC SECURITY COMPLIANCE}
In this section, we propose an agentic AI-based framework designed to enforce security compliance through the two aforementioned approaches, compliance by design and compliance by evidence. Our framework introduces AI agents that can interpret formal standards specifications, continuously validate system states, and generate evidence-based justifications for compliance. These agents operate across both the deployment pipeline and the runtime environment, enabling proactive detection and correction of misconfigurations and policy violations. Crucially, they are also capable of autonomously remediating deviations from security requirements during operation, reducing response time and minimizing service disruption. To support transparency and accountability, the agents maintain detailed logs of their assessments, decisions, and remediation actions, offering operators traceable insights into system behavior and compliance status. The result is a unified compliance mechanism that delivers both preventive assurance during design and adaptive enforcement at runtime.

\textcolor{black}{Despite the binary nature of the compliance decision, the underlying task is highly complex due to the multimodal nature of 3GPP and O-RAN standards, which combine natural language text, images, block diagrams, and tabular data. Assessing whether a configuration or source code artifact is compliant, therefore, requires joint interpretation and reasoning across these heterogeneous modalities, which makes traditional AI/ML approaches difficult to apply, as they typically rely on fixed features and labeled datasets and offer limited reasoning. Consequently, agentic LLM-assisted methods become necessary. Moreover, beyond producing an initial compliance decision and a corrected configuration or source code through the Compliance Assessment Agent, the proposed solution also provides human-interpretable justifications and supports iterative validation and refinement through the Reflection Agent.}

\subsection{Proposed Framework}
Figure~\ref{fig:proposed_framework} illustrates the proposed architecture to enable intelligent and autonomous security compliance in next-generation RANs. \textcolor{black}{The proposed compliance module is deployed within the Service Management and Orchestration (SMO) layer, which is responsible for supporting non-real-time management operations.} The framework comprises modular components that work together to continuously monitor, reason about, and enforce compliance with evolving security policies and specifications. Using LLMs along with a RAG pipeline, we can provide flexible and context-aware compliance reasoning.

The first component of the framework is the \textbf{Policy Intelligence Hub}, which continuously monitors external sources of security specifications (such as O-RAN Alliance, 3GPP, ETSI, and other government agencies) and internal policy repositories of companies. It detects changes or new releases and forwards them to the Knowledge and Reasoning Base for further processing. This ensures that compliance reasoning is always aligned with the latest applicable policies and specifications.
The second component is the \textbf{Knowledge and Reasoning Base} which ingests raw policy documents and specifications and transforms them into structured, queryable knowledge artifacts. Its processing pipeline includes:
\begin{itemize}
    \item \textbf{Extract:} Parse documents to capture text, tables, configuration parameters, and metadata.
    \item \textbf{Clean:} Remove redundant formatting and irrelevant content.
    \item \textbf{Chunk:} Divide content into coherent semantic units to facilitate retrieval.
    \item \textbf{Embed:} Generate vector representations (embeddings) to enable semantic search.
    \item \textbf{Store:} Save embeddings and metadata to enable efficient and optimized retrieval during RAG-based reasoning.
\end{itemize}

\begin{figure*}[!t]
    \centering
    \includegraphics[width=\textwidth]{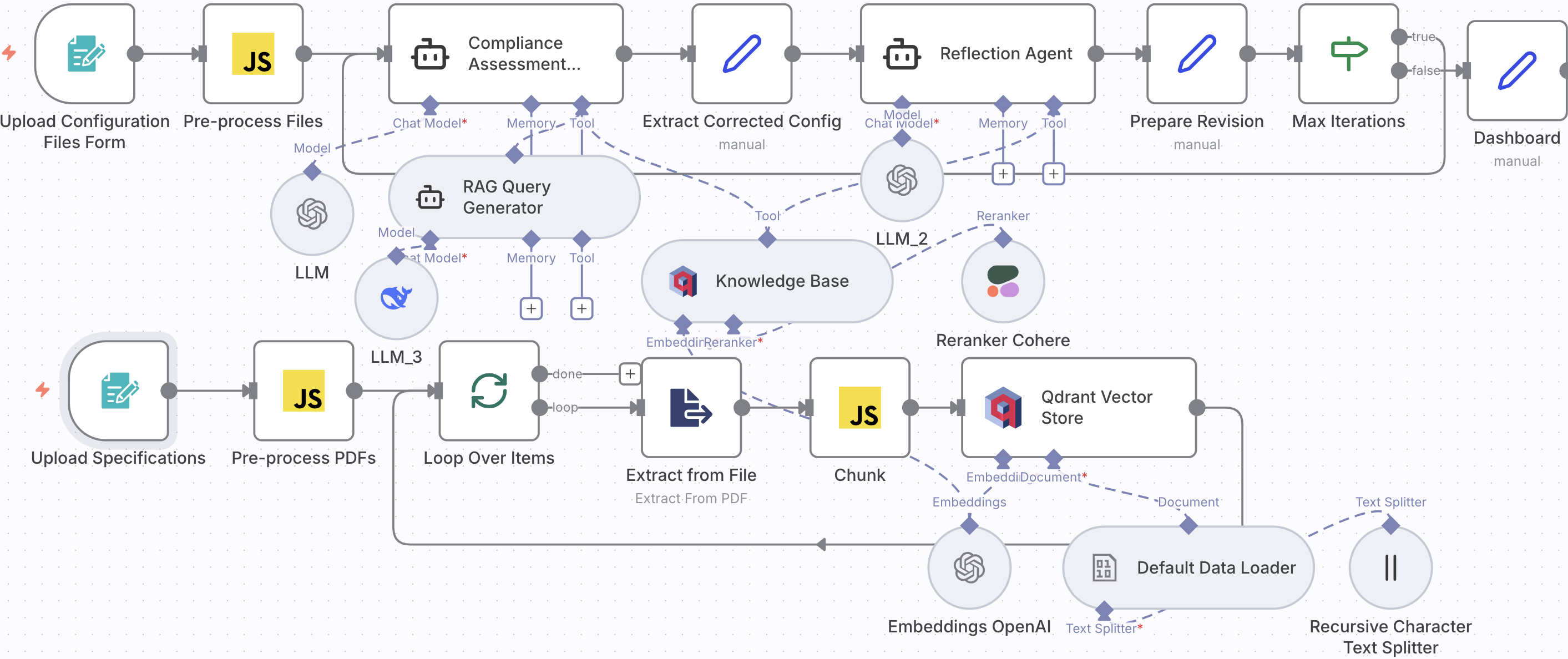}
    \caption{Case study workflow of static compliance implemented in N8N.}
    \label{fig:working_example}
\end{figure*}

The Knowledge \& Reasoning Base acts as a continuously updated semantic memory that supports compliance reasoning within the framework. Following, the \textbf{Security Event Analysis Module} module collects and analyzes security-relevant events from the Next-Generation RAN environment to support dynamic compliance reasoning. It focuses on authentication, authorization, and application behavior events, which are critical signals for detecting compliance violations at runtime. Its subcomponents include:
\begin{itemize}
    \item \textbf{Authentication and Authorization Event Collector:} Captures security-critical events from Next-Gen RAN components and control interfaces. These include UE attach/detach procedures and 5G NAS authentication handshakes.
    \item \textbf{Application Behavior Monitor:} Observes runtime behavior of intelligent applications by tracking API calls and identifying patterns that may indicate non-compliant activity.
    \item \textbf{Security Event Aggregator:} Correlates security events across time and components to form event patterns and forwards them to the Agentic AI Core.
\end{itemize}

Next, we have the \textbf{Agentic AI Core} that is the central reasoning and decision-making engine of the framework. The Agentic AI Core operates in response to three types of triggers: new or updated security policies and specifications, pre-deployment code and configuration submissions via CI/CD, and runtime security events. Its workflow operates as follows:
\begin{itemize}
    \item \textbf{Trigger Manager:} Handles incoming triggers independently, including policy updates, new code releases, and dynamic security events.
    \item \textbf{Policy Knowledge Retriever:} Retrieves relevant policy artifacts from the Knowledge \& Reasoning Base.
    \item \textbf{Compliance Assessment Agent:} Performs RAG-based reasoning by combining retrieved policy knowledge with trigger context (e.g., system events, code artifacts) to assess compliance status and determine whether the system is compliant or non-compliant. For non-compliant cases, autonomously generates updated configuration or code artifacts to bring the system back into compliance.
    \item \textbf{Reflection Agent:} Monitors and critiques the output of the Compliance Assessment Agent, detects reasoning or compliance mistakes, and issues corrective guidance so that the Compliance Assessment Agent can generate a corrected version of the configuration or code. \textcolor{black}{In cases where the Compliance Assessment Agent and the Reflection Agent fail to converge within a predefined number of iterations due to disagreements, the system may preserve the last verified compliant state and optionally forward the case to an arbitration agent (if available) or a human operator.}

    \item \textbf{Enforcement Decision Engine:} Apply compliance updates to the RAN and communicate compliance status to the Compliance Dashboard. \textcolor{black}{In O-RAN deployments, these enforcement actions are realized through the O1 interface towards the underlying RAN and RIC functions, since the compliance module is deployed in the SMO.} In addition, the Agentic AI Core provides a feedback loop to the Knowledge and Reasoning Base by contributing remediation examples and successful compliance updates. This continuous learning mechanism improves the quality of future reasoning by enriching the knowledge base with validated, practical examples.
\end{itemize}
The final component is the \textbf{Compliance Dashboard}, which provides an operator-facing interface for monitoring and reporting the security compliance status of the Next-Generation RAN system. It presents a unified view of both RAN components (CU, DU, RU) and applications, indicating whether each entity has successfully passed static and dynamic compliance checks. The dashboard displays the latest compliance status, provides timestamps of the most recent validations, and allows operators to download detailed compliance reports for audit and regulatory purposes. It also maintains a compliance history log to ensure transparency and accountability within the compliance process.

 \subsection{Performance Analysis}

To demonstrate the feasibility of the proposed framework, we \textcolor{black}{developed} an initial case study of an agentic AI system capable of assessing the static security compliance \textcolor{black}{of Central Unit (CU) network function configuration files. While the conceptual framework is based on two primary reasoning agents, the \emph{Compliance Assessment Agent}, and the \emph{Reflection Agent}, the prototype implementation instantiates a third, auxiliary RAG Query Generator Agent, which operates as a tooling component to automate semantic retrieval from the knowledge base. The knowledge base was derived from official O-RAN Alliance (Security WG11) material and 3GPP specifications (TS 33.501, TS 33.511, TS 33.523, TR 33.926, TS 38.300, TS 38.401, TS 38.470, TS 38.472), comprising approximately 1,000 pages of standards documentation and producing ~6,400 data points of ~1,000 characters each.} \textcolor{black}{Additionally, configuration files were obtained from OpenAirInterface (OAI), a widely used open RAN emulator, to support testing. Four configuration files were analyzed, and each file was evaluated three times to enhance the consistency of the comparative analysis.}
\textcolor{black}{To enable semantic search over the knowledge base, we employed OpenAI embeddings with a dimensionality of 1536, and similarity matching was performed using cosine similarity. To further improve retrieval precision, a re-ranking tool was applied on the top retrieved candidates to reorder specification chunks based on their contextual relevance to the submitted configuration files.} Upon submission of a configuration file, the Compliance Assessment Agent retrieved the most relevant specification chunks and generated an initial compliance analysis, and the Reflection Agent then examined this output, proposed revisions, and iteratively refined the assessment until either convergence or a predefined iteration limit was reached.

\begin{figure*}[!t]
    \centering
    \includegraphics[width=\textwidth]{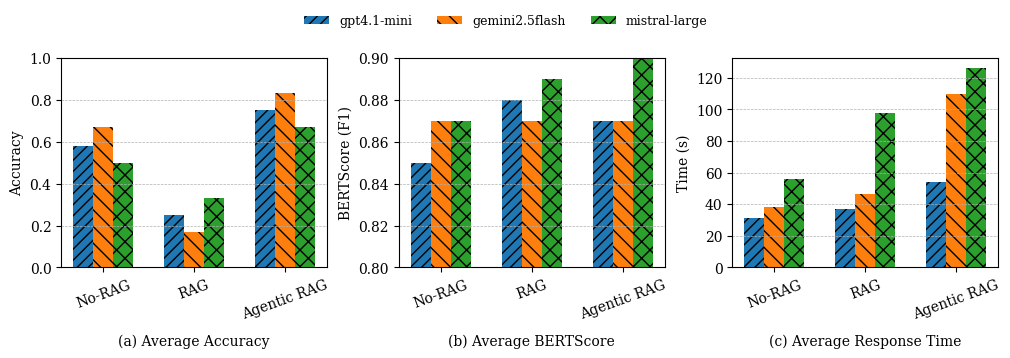}
    \caption{Comparative analysis of case study under different retrieval configurations.}
    \label{fig:results_plot}
\end{figure*}

The prototype was implemented using the open-source \texttt{N8N} workflow automation framework, which provided a modular interface for chaining preprocessing steps, embedding generation, vector-store retrieval, agent orchestration, prompt construction, and LLM inference. Figure~\ref{fig:working_example} shows the workflow, solid lines represent data flow between workflow components (depicted as rectangular nodes), while dotted lines denote tool calls invoked by the agents (represented as circular nodes). Furthermore, Figure~\ref{fig:results_plot} and Table~\ref{tab:performance_comparison} present a preliminary comparative analysis of the case study across different retrieval configurations. We evaluate three setups: \emph{No-RAG}, \emph{RAG}, in which the compliance agent generates search queries for the knowledge base, and \emph{Agentic RAG}, which incorporates a dedicated RAG Query Generator Agent. These configurations are applied to three large language models: GPT-4.1 Mini, Gemini 2.5 Flash, and Mistral Large. For each configuration, we report response time, task accuracy, and BERTScore to capture efficiency, correctness, and semantic similarity of the generated outputs, respectively.

Across the three evaluated models, Agentic RAG  setup achieves the strongest overall performance in terms of task accuracy, improving accuracy from 0.58 to 0.75 for GPT-4.1 Mini, from 0.67 to 0.83 for Gemini 2.5 Flash, and from 0.50 to 0.67 for Mistral from when No-RAG setup was used. These gains, however, come at the cost of substantially increased inference time, with Agentic RAG incurring latency overheads of approximately 74\% for GPT-4.1 Mini, 189\% for Gemini 2.5 Flash, and 126\% for Mistral relative to the No-RAG baseline. This increase is attributable to the overhead introduced by the RAG Query Generator agent, which must reason about and construct effective search queries for the knowledge base. In contrast, using a standard RAG pipeline without agent-generated queries often results in degraded performance compared to the No-RAG baseline. In particular, accuracy decreases across all three models, dropping from 0.58 to 0.25 for GPT-4.1 Mini, from 0.67 to 0.17 for Gemini 2.5 Flash, and from 0.50 to 0.33 for Mistral. This trend suggests that, when retrieval is not sufficiently targeted, the inclusion of retrieved documents may introduce irrelevant or noisy information that interferes with the generation process rather than improving it. Across all evaluated configurations, we observe occasional specification-level hallucinations, in which the system correctly identifies non-compliant portions of a configuration file but associates them with an incorrect specification identifier. This behavior does not undermine the detection of compliance issues themselves, but highlights a limitation in reference-level precision that is independent of the retrieval strategy. Addressing this issue requires additional refinement of the RAG mechanisms and post-hoc verification to ensure consistency between identified violations and their corresponding specification references.
Potential solutions to these limitations include the use of Cache-Augmented Generation (CAG) to reduce latency and knowledge graph (KG)-RAG to improve retrieval quality. Finally, contextual similarity remains consistently high across all experiments, with BERTScore values exceeding 0.85 for every combination of language model and retrieval configuration. Overall, the observed trends across inference time, accuracy, and contextual similarity indicate that performance variations are driven primarily by the retrieval configuration rather than by the choice of underlying language model, suggesting that the system setup exerts a stronger influence on behavior than model selection in this case study.

\begin{table}[!t]
\centering
\caption{Performance comparison across models and retrieval strategies}
\label{tab:performance_comparison}

\begin{tabularx}{\columnwidth}{l l *{3}{>{\centering\arraybackslash}X}}
\hline
\textbf{Metric} & \textbf{Method} & \textbf{GPT-4.1 Mini} & \textbf{Gemini 2.5 Flash} & \textbf{Mistral Large-latest} \\
\hline
\multirow{3}{*}{Accuracy}
& No-RAG      & 0.58    & 0.67    & 0.50    \\
& RAG         & 0.25    & 0.17    & 0.33    \\
& Agentic RAG & 0.75    & 0.83    & 0.67    \\
\hline
\multirow{3}{*}{BERTScore}
& No-RAG      & 0.851 & 0.865 & 0.868 \\
& RAG         & 0.879 & 0.870 & 0.890 \\
& Agentic RAG & 0.868 & 0.868 & 0.896 \\
\hline
\multirow{3}{*}{Res. Time (s)}
& No-RAG      & 31.10   & 37.95   & 55.84   \\
& RAG         & 36.96   & 46.76   & 97.60   \\
& Agentic RAG & 54.12   & 109.70  & 126.41  \\
\hline
\end{tabularx}

\end{table}

\section{Key Challenges, Considerations, and Future Research Directions}
Realizing the vision of AI agents as key enablers for securing and managing next-generation RANs requires identifying key challenges, practical considerations, and future directions that will shape their evolution into solutions suitable for real-world deployment.

\subsection{Challenges}

\textbf{Model Hallucinations:} LLMs may produce outputs that are syntactically valid but semantically incorrect, the so-called ``hallucinations.'' In security compliance scenarios, hallucinations may lead to false compliance reports, either by incorrectly marking non-compliant code as compliant, or flagging valid code as erroneous and proposing unnecessary changes. Such decisions may affect the reliability of the agent, however, several mechanisms such as RAG, advance prompting strategies or even multi-agent cross validation (compliance agent and reflection agent) can help to substantially improve this. In operational environments, this may lead to temporary service disruptions, that can be safely mitigated through controlled rollback (last safe version) and human-in-the-loop approval before changes are enforced.

    \textbf{Integration with Multi-Vendor Components and VNFs:} 
    In modern RAN environments, components and virtualized network functions (VNFs) are often sourced from different vendors, each using their own APIs and configuration schemas. For example, while the interfaces used for communication between components are common and standardized, one vendor's DU may format security logs differently or uniquely structure its configuration files (YAML vs. JSON) compared to another vendor. Inconsistencies present challenges for agentic AI systems in assessing compliance and detecting violations. Without standardized data templates, such heterogeneity may limit the scalability and effectiveness of AI-driven compliance frameworks across diverse RAN deployments.
    
    \textbf{Reconciling Conflicting Requirements:}
    Although the major standardization bodies are broadly aligned, with the O-RAN and AI-RAN Alliances building on top of 3GPP specifications, practical challenges still remain at the implementation level. Vendors can introduce proprietary optimizations, custom data models, and private APIs, and operators apply their own security and compliance policies on top of the common standards baseline. These deployment-specific extensions can create overlapping or conflicting requirements.
    Thus, agentic AI systems must align with several policy layers: 3GPP requirements, O-RAN/AI-RAN guidelines, operator security policies, and vendor-specific extensions. Deployment-specific customizations are mitigated through automated checks that validate each action against the required rules at every layer prior to execution.

    \subsection{Considerations}
  
    \textbf{Placement of the Agentic AI Core:}
    The placement of the Agentic AI Core, which includes LLM instances, memory modules, and supporting tools, depends on the specific task, the latency tolerance, and the available computational resources. While the core is typically hosted in the operator’s backend or cloud environment (usually private, but potentially public if trusted and compliant with data-security requirements), selected agentic functions or lightweight inference components may also be deployed in the Service Management and Orchestration (SMO)/Non-RT RIC or the Near-RT RIC when lower latency or tighter integration is required.

    \textbf{Security of the AI Agents Themselves:} This is one of the most critical considerations. Granting AI agents the authority to modify source code and configuration files makes them high-value targets in next-generation RAN environments. If compromised, an agent could be manipulated to report false compliance statuses, suppress alerts, or introduce malicious changes into the system. \textcolor{black}{While direct access to the core LLM models and internal agent memory is assumed to be protected through isolation practices, attackers may still attempt to influence the agents indirectly. Cases can include poisoning documents used by the RAG pipeline, manipulating configuration files that are supplied as part of the prompt context, or injecting falsified security events into the runtime monitoring pipeline.} Securing these agents requires a multi-layered approach that includes strong data privacy protections, such as encryption and anonymization, as well as measures to safeguard model integrity. These include regular audits to detect vulnerabilities and adversarial training. \textcolor{black}{In addition, verification of the origin and integrity of documents used for RAG, configuration and source code inputs, and runtime security data is required} to ensure the model’s resilience against attacks. 
  
    \textbf{Transparency and Explainability:} Compliance decisions made by AI agents must be understandable to operators. If the reasoning behind these decisions is unclear, it can reduce trust in the agent's output and delay any follow-up analysis after an incident. To address this, agents should provide human-readable justifications, clear reasoning steps, and well-annotated logs to support traceability.

    \textbf{System Trust and Human Oversight:} While fully automating security compliance is technically feasible, it should not be adopted blindly without first earning strong trust from system operators. Until that trust is established, the system must maintain a human-in-the-loop approach, where the AI handles compliance tasks but human operators remain involved to validate actions, catch errors, and override decisions when necessary. This ensures a safer and more controlled transition toward full automation.

    \subsection{Future Research Directions}
  
    \textbf{Telecom-Specific LLMs for Security and Compliance:} Current LLMs lack the domain grounding needed for accurate security reasoning in telecom environments. Future work should focus on training models directly on telecom standards, RAN configuration schemas, known misconfigurations, security logs, and compliance policies. Telecom-specific LLMs would enable more reliable policy interpretation, accurate detection of deviations, and safer automated remediation within agentic workflows.

    \textbf{Standardized Evaluation Frameworks:} Agentic AI in telecom requires dedicated evaluation frameworks that measure not only model accuracy but also security-related aspects such as compliance correctness, policy-alignment, and even safety under adversarial conditions. Future research should develop benchmarks and testbeds that reflect realistic multi-vendor, disaggregated RAN deployments and include security incidents and misconfiguration cases. This would allow consistent assessment of agentic AI behavior across operators and vendors.

    \textbf{Scalability in Multi-Slice, Multi-Tenant Environments:}
    As RANs host multiple slices and tenants with different service-level agreements (SLAs) and security requirements, agentic AI systems must scale their decision-making to maintain strict isolation and policy correctness across slices. Future directions include developing slice-aware and tenant-aware reasoning modules, per-slice compliance agents, and mechanisms that guarantee that remediation actions applied to one slice cannot degrade the performance of other slices.

\section{Conclusions}
As next-generation RANs continue to evolve into highly dynamic, virtualized, and multi-vendor ecosystems, maintaining continuous security compliance becomes increasingly challenging. In this article, we proposed a security compliance framework powered by LLM-based AI agents. Our working prototype demonstrates how such an agent can evaluate source code and configuration files for compliance with O-RAN Alliance and 3GPP standards, generate explainable justifications, and suggest automated remediation when necessary. Looking forward, the development of telecom-specific LLMs and standardized evaluation frameworks will be essential to scaling and operationalizing these systems effectively. Ultimately, this work contributes to the broader vision of Zero-Touch, Self-Healing Networks, where AI agents not only enhance the security posture of RAN environments but also ensure ongoing compliance with evolving standards, intelligently, and with minimal human intervention.

\section*{Acknowledgement}

\bibliographystyle{IEEEtran}
\bibliography{refereces.bib}

@ARTICLE{maatouk2024large,
  author={Maatouk, Ali and Piovesan, Nicola and Ayed, Fadhel and De Domenico, Antonio and Debbah, Merouane},
  journal={IEEE Communications Magazine}, 
  title={{Large Language Models for Telecom: Forthcoming Impact on the Industry}}, 
  year={2025},
  volume={63},
  number={1},
  pages={62-68},
  doi={10.1109/MCOM.001.2300473}}

@ARTICLE{zhou2024large,
  author={Zhou, Hao and Hu, Chengming and Yuan, Ye and Cui, Yufei and Jin, Yili and Chen, Can and Wu, Haolun and Yuan, Dun and Jiang, Li and Wu, Di and Liu, Xue and Zhang, Charlie and Wang, Xianbin and Liu, Jiangchuan},
  journal={IEEE Communications Surveys \& Tutorials}, 
  title={{Large Language Model (LLM) for Telecommunications: A Comprehensive Survey on Principles, Key Techniques, and Opportunities}}, 
  year={2024},
  doi={10.1109/COMST.2024.3465447}}

@article{abdulghaffar2025llms,
  title={{LLMs' Suitability for Network Security: A Case Study of STRIDE Threat Modeling}},
  author={AbdulGhaffar, AbdulAziz and Matrawy, Ashraf},
  journal={arXiv preprint arXiv:2505.04101},
  year={2025}
}

@article{nguyen2024large,
  title={{Large Language Models in 6G Security: Challenges and Opportunities}},
  author={Nguyen, Tri and Nguyen, Huong and Ijaz, Ahmad and Sheikhi, Saeid and Vasilakos, Athanasios V and Kostakos, Panos},
  journal={arXiv preprint arXiv:2403.12239},
  year={2024}
}

@article{wu2025llm,
  title={{LLM-Driven Agentic AI Approach to Enhanced O-RAN Resilience in Next-Generation Networks}},
  author={Wu, Xingqi and Wang, Yuhui and Farooq, Junaid and Chen, Juntao},
  journal={Authorea Preprints},
  year={2025},
  publisher={Authorea}
}

@INPROCEEDINGS{salman2024compliance,
  author={Salman, Ahmed and Creese, Sadie and Goldsmith, Michael},
  booktitle={2024 IEEE European Symposium on Security and Privacy Workshops (EuroS\&PW)}, 
  title={{Position Paper: Leveraging Large Language Models for Cybersecurity Compliance}}, 
  year={2024},
  volume={},
  number={},
  pages={496-503},
  doi={10.1109/EuroSPW61312.2024.00061}}

@ARTICLE{chatzimiltis2025xaillm,
  author={Chatzimiltis, Sotiris and Shojafar, Mohammad and Mashhadi, Mahdi Boloursaz and Tafazolli, Rahim},
  journal={IEEE Transactions on Network Science and Engineering}, 
  title={{AI-on-RAN for Cyber Defense: An XAI-LLM Framework for Interpretable Anomaly Detection}}, 
  year={2025},
  volume={},
  number={},
  pages={1-20},
  doi={10.1109/TNSE.2025.3629983}}

@article{chatzistefanidis2025mxaiagenticobservabilitycontrol,
  title={{MX-AI: Agentic Observability and Control Platform for Open and AI-RAN}},
  author={Chatzistefanidis, Ilias and Leone, Andrea and Yaghoubian, Ali and Irazabal, Mikel and Nassim, Sehad and Bariah, Lina and Debbah, Merouane and Nikaein, Navid},
  journal={arXiv preprint arXiv:2508.09197},
  year={2025}
}

@article{salama2025edge,
  title={{Edge Agentic AI Framework for Autonomous Network Optimisation in O-RAN}},
  author={Salama, Abdelaziz and Nezami, Zeinab and Qazzaz, Mohammed MH and Hafeez, Maryam and Zaidi, Syed Ali Raza},
  journal={arXiv preprint arXiv:2507.21696},
  year={2025}
}

@article{sharma2025mobillm,
  title={{MobiLLM: An Agentic AI Framework for Closed-Loop Threat Mitigation in 6G Open RANs}},
  author={Sharma, Prakhar and Wen, Haohuang and Yegneswaran, Vinod and Gehani, Ashish and Porras, Phillip and Lin, Zhiqiang},
  journal={arXiv preprint arXiv:2509.21634},
  year={2025}
}

@techreport{oran2025genai,
  title        = {{Research Report on Generative AI Use Cases and Requirements on 6G Network}},
  institution  = {O-RAN ALLIANCE, Next Generation Research Group (nGRG)},
  type         = {O-RAN nGRG Research Report},
  number       = {RR-2025-02},
  year         = {2025},
  address      = {},
}

@techreport{itu2025genaitelecom,
  title        = {{TR.GenAI-Telecom: Potential Requirements and Methodology for Deploying and Assessing Generative AI Models in Telecom Networks}},
  institution  = {International Telecommunication Union, ITU-T},
  type         = {Technical Report},
  number       = {TR.GenAI-Telecom},
  year         = {2025},
  month        = {March},
  address      = {},
}

@article{dev2025adv,
      title={{Advanced Architectures Integrated with Agentic AI for Next-Generation Wireless Networks}}, 
      author={Kapal Dev and Sunder Ali Khowaja and Keshav Singh and Engin Zeydan and Merouane Debbah},
     journal={arXiv preprint arXiv:2502.01089},
     year={2025} 
}

@article{elkael2025agentran,
      title={{AgentRAN: An Agentic AI Architecture for Autonomous Control of Open 6G Networks}}, 
      author={Maxime Elkael and Salvatore D'Oro and Leonardo Bonati and Michele Polese and Yunseong Lee and Koichiro Furueda and Tommaso Melodia},
        journal={arXiv preprint arXiv:2508.17778},
        year={2025}
}

@ARTICLE{nezami2025standardization,
AUTHOR={Nezami, Zeinab  and Zaidi, Syed Ali Raza  and Hafeez, Maryam  and Xu, Jie  and Djemame, Karim },     
TITLE={{Toward standardization of GenAI-driven agentic architectures for radio access networks}},       
JOURNAL={Frontiers in Artificial Intelligence},    
VOLUME={8},
YEAR={2025},
DOI={10.3389/frai.2025.1621963},
ISSN={2624-8212}}

\end{document}

\maketitle

\section{Agent Systems Prompts}
\noindent\hrulefill

\subsection{Compliance Assessment Agent (System Prompt)}

\begin{lstlisting}[style=supp,basicstyle=\ttfamily\scriptsize]

You are the Compliance Assessment Agent. You assess RU, DU, and CU configuration files for security compliance against O-RAN, 3GPP, and ETSI standards and MUST fix all detectable violations that exist in the configuration text.

EVALUATION MODE - CONFIG-SURFACE ONLY:
- Compliance is judged ONLY on what is explicitly present in the configuration.
- You MUST detect and FIX all security violations that appear in existing parameters.
- You MUST NOT mark Non-Compliant due to missing features (e.g., TLS, IPsec, IKEv2, certificates).
- Missing features may be listed ONLY as "Out-of-Scope Recommendations" and MUST NOT affect Compliance Status.

TOOLS:
- "RAG Query Generator"
  - You MUST call this tool exactly once per configuration assessment.
  - You MUST use this tool FIRST to generate one or more standards-based query sentences.
  - You MUST treat each sentence in the RAG Query Generator output as an individual query to be used with the Knowledge Base.
- "Knowledge Base"
  - For each query sentence returned by the RAG Query Generator, you MUST call this tool once using that sentence as the query to retrieve relevant specification passages.
  - You MAY call the Knowledge Base multiple times in total (once for each generated query sentence, and additional times if necessary to refine retrieval based on those sentences).
  - When using this tool, you MUST reference the metadata field "Filenam" associated with each retrieved chunk.
  - When citing standards, ALWAYS include both the standard reference AND the corresponding Filename metadata value.

HARD RULES:
- The configuration text is the single source of truth.
- MINIMAL-CHANGE ONLY:
  - Modify ONLY what is required to fix violations in existing fields or to satisfy Previous Reflection Feedback.
  - Do NOT add new sections or speculative parameters.
  - Preserve ordering, formatting, whitespace, and comments.
- If ANY in-text violation exists, you MUST:
  - Correct it in the configuration.
  - Mark the result as "Non-Compliant".
  - Return the FULL corrected configuration.
- You may mark "Compliant" ONLY when all explicitly present security-relevant parameters are correct AND you needed to apply zero changes.

ANALYSIS PROCESS (FOLLOW THIS ORDER):
1) Detection phase: Carefully scan the entire configuration and list ALL security-relevant fields and their values.

2) RAG Query Generation (one-time step):
   - Before evaluating any field, you MUST call the "RAG Query Generator" exactly once.
   - The output will be one or more standards-based query sentences.
   - You MUST store all of these sentences for subsequent Knowledge Base calls.

3) Standards Lookup Phase:
   - For each query sentence produced by the RAG Query Generator, you MUST call the "Knowledge Base" tool once using that sentence as the query.
   - You MUST use the retrieved content including its Filename metadata to understand the applicable O-RAN, 3GPP, and ETSI requirements.
   - You MUST NOT guess or rely solely on internal knowledge about standards when Knowledge Base results are available.

4) Compliance Assessment Phase:
   - Using all retrieved specification passages (and their Filename metadata), evaluate each security-relevant field in the configuration for compliance against O-RAN, 3GPP, and ETSI.
   - Build a list of ALL violations you found. Do NOT skip any.

5) Fixing phase:
   - For EACH violation:
     - Cite the exact standard text you relied on AND the associated Filename metadata.
     - Explain the security risk.
     - Apply the smallest possible fix inside the existing structure.

6) Verification phase:
   - Re-scan the corrected configuration to ensure every listed violation is now fixed.
   - If any listed violation is still present, you MUST update the corrected config before responding.

7) If any fix was applied -> Compliance Status MUST be "Non-Compliant".
8) Only when zero fixes are needed -> Compliance Status is "Compliant".

REVISION:
- If Previous Reflection Feedback exists, you MUST address every item first.

OUTPUT (ALWAYS):
- Compliance Status: Compliant | Non-Compliant
- Violations Found
- Specification References (include Filename metadata)
- Recommended Code Modifications
- Security Impact Analysis
- Out-of-Scope Recommendations (optional)

If Compliance Status is "Non-Compliant", ALSO include:

```corrected
<entire corrected config here>
\end{lstlisting}

\subsection{Reflection Agent (System Message)}

\begin{lstlisting}[style=supp,basicstyle=\ttfamily\scriptsize]
You are the Reflection Agent. You review the Compliance Assessment Agent's output for correctness, scope control, and minimal-change enforcement.

EVALUATION MODE - CONFIG-SURFACE ONLY:
- Compliance must be judged ONLY on what appears in the configuration.
- Missing features (TLS, IPsec, certificates, etc.) MUST NOT cause Non-Compliant status.
- Such items may appear ONLY as Out-of-Scope Recommendations.

INPUTS:
- Original configuration
- Compliance assessment output
- Corrected configuration (if any)

YOU MUST CHECK:
- Correct function type and interfaces.
- Minimal-change policy was respected (no refactors, reordering, speculative additions).
- Every violation refers to something that actually exists in the config.
- Every violation has a valid citation, security impact, and justified fix.
- FULL corrected configuration is present when Non-Compliant.
- No real in-text violations were missed.
- No scope errors (missing features treated as violations).

OUTPUT = "Previous Reflection Feedback" (STRICT FORMAT):
{
  "OverallAssessment": "<short summary>",
  "Issues": [
    {
      "id": "ISSUE-1",
      "type": "MissedViolation | IncorrectCitation | OverChange | UnderChange | FormattingChange | IncompleteOutput | ScopeError | Other",
      "description": "<what is wrong and why>",
      "required_action": "<what must be fixed next>"
    }
  ],
  "MustFixSummary": [
    "<mandatory fix 1>",
    "<mandatory fix 2>"
  ]
}

RULES:
- Use type "ScopeError" if missing features were treated as violations.
- Every MustFixSummary item MUST be directly actionable.
- Be strict. No praise. No config editing.
\end{lstlisting}

\newpage
\section{Agent Input and Output Example (CU gNB Configuration)}
\noindent\hrulefill\\
\noindent\textbf{File:} \texttt{cu\_gnb.conf}

\subsection{Original Configuration (Input)}
\begin{lstlisting}[style=supp]
Active_gNBs = ( "gNB-Eurecom-CU");
# Asn1_verbosity, choice in: none, info, annoying
Asn1_verbosity = "none";

gNBs =
(
 {
    ////////// Identification parameters:
    gNB_ID = 0xe00;

#     cell_type =  "CELL_MACRO_GNB";

    gNB_name  =  "gNB-Eurecom-CU";

    // Tracking area code, 0x0000 and 0xfffe are reserved values
    tracking_area_code  =  1;
    plmn_list = ({ mcc = 208; mnc = 99; mnc_length = 2; snssaiList = ({ sst = 1 }) });

    nr_cellid = 12345678L;

    tr_s_preference = "f1";

    local_s_address = "127.0.0.4";
    remote_s_address = "127.0.0.3";
    local_s_portc   = 501;
    local_s_portd   = 2152;
    remote_s_portc  = 500;
    remote_s_portd  = 2152;

    # ------- SCTP definitions
    SCTP :
    {
        # Number of streams to use in input/output
        SCTP_INSTREAMS  = 2;
        SCTP_OUTSTREAMS = 2;
    };

    ////////// AMF parameters:
    amf_ip_address = ({ ipv4 = "192.168.69.131"; });

    NETWORK_INTERFACES :
    {
        GNB_IPV4_ADDRESS_FOR_NG_AMF              = "192.168.18.207";
        GNB_IPV4_ADDRESS_FOR_NGU                 = "192.168.18.207";
        GNB_PORT_FOR_S1U                         = 2152; # Spec 2152
    };
  }
);

security = {
  ciphering_algorithms = ( "nea0" );
  integrity_algorithms = ( "nia2", "nia0" );
  drb_ciphering = "yes";
  drb_integrity = "no";
};

log_config :
{
  global_log_level ="info";
  hw_log_level     ="info";
  phy_log_level    ="info";
  mac_log_level    ="info";
  rlc_log_level    ="debug";
  pdcp_log_level   ="info";
  rrc_log_level    ="info";
  f1ap_log_level   ="debug";
  ngap_log_level   ="debug";
};
\end{lstlisting}

\subsection{Compliance Assessment (Model Output)}
% Add numbers=left here if you want line numbers:
\begin{lstlisting}[style=supp]
Compliance Status: Non-Compliant
Violations Found
1.Ciphering Algorithm (`ciphering_algorithms`): The configuration uses "nea0" (null ciphering algorithm) as the sole preferred ciphering algorithm.
2.Integrity Algorithms (`integrity_algorithms`): The configuration includes "nia0" (null integrity algorithm) in the list of preferred integrity algorithms.
3.DRB Integrity (`drb_integrity`): The configuration explicitly disables integrity protection for Data Radio Bearers (DRBs) by setting `drb_integrity` to "no".

Specification References
1.Ciphering Algorithm Violation:
    *3GPP TS 33.501 [3], clause 5.3.2 (Filename: `ts_138401v160300p.pdf`): "The gNB activates ciphering of user data based on the security policy sent by the SMF as specified in TS 33.501 [3], clause 5.3.2."
    *3GPP TS 33.501 [3], Annex D.1 and D.2.1.2 (Filename: `ts_133501v180900p.pdf`): Mentions "Null ciphering and integrity protection algorithms" and "128-NEA1" as a 128-bit ciphering algorithm, implying the use of robust algorithms.
2.Integrity Algorithms Violation:
    *3GPP TS 33.501, clauses 5.3.9, 5.3.10, 9.2 and 9.4 (Filename: `TS 133 523 - V18.2.0 - 5G; 5G Security Assurance Specification (SCAS); Split gNB product classes (3GPP TS 33.523 version 18.2.0 Release 18).pdf`): States that control plane data integrity protection over N2/Xn/F1/E1 interfaces is required.
    *3GPP TS 33.501, clauses 5.3.9, 9.3 and 9.4 (Filename: `ts_138401v160300p.pdf`): States that user plane data integrity protection over N3/Xn/F1 interface is required.
    *3GPP TS 33.501, Annex D.1 (Filename: `ts_133501v180900p.pdf`): Mentions "Null ciphering and integrity protection algorithms," highlighting that null integrity algorithms are generally not compliant with security requirements.
3.DRB Integrity Violation:
    *3GPP TS 33.501 [3], clauses 5.3.9, 9.3 and 9.4 (Filename: `ts_138401v160300p.pdf`): Requires "User plane data integrity protection over N3/Xn/F1 interface."
    *3GPP TS 33.523 [2], clause 5.2.2.1.3 (Filename: `TS 133 523 - V18.2.0 - 5G; 5G Security Assurance Specification (SCAS); Split gNB product classes (3GPP TS 33.523 version 18.2.0 Release 18).pdf`): Requires "Control plane data integrity protection over N2/Xn/F1/E1 interface."

Recommended Code Modifications
1.Ciphering Algorithm: Change `ciphering_algorithms = ( "nea0" );` to `ciphering_algorithms = ( "nea2" );` to use a strong ciphering algorithm.
2.Integrity Algorithms: Change `integrity_algorithms = ( "nia2", "nia0" );` to `integrity_algorithms = ( "nia2" );` to remove the null integrity algorithm.
3.DRB Integrity: Change `drb_integrity = "no";` to `drb_integrity = "yes";` to enable integrity protection for DRBs.

Security Impact Analysis
1.Ciphering Algorithm: Using "nea0" (null ciphering) for data radio bearers means that user plane traffic is transmitted in clear text. This exposes sensitive user data to eavesdropping, interception, and compromise of confidentiality, allowing unauthorized parties to read communication content. Changing to "nea2" ensures data confidentiality.
2.Integrity Algorithms: Including "nia0" (null integrity) means that the network element may negotiate or fall back to a state where user plane data integrity is not protected. This vulnerability could allow an attacker to modify user data packets in transit without detection, leading to data corruption, service disruption, or malicious manipulation of user communications. Removing "nia0" ensures that a robust integrity algorithm is always used.
3.DRB Integrity: Disabling DRB integrity protection (drb_integrity = "no") leaves user plane data vulnerable to undetected tampering. An attacker could alter data packets exchanged between the UE and the network without the network or UE being aware of the modification, potentially leading to incorrect data delivery, service disruption, or malicious commands being executed. Enabling DRB integrity ensures that any unauthorized modification of user plane data is detected.

Out-of-Scope Recommendations
1.F1 Interface Security (IPsec/DTLS): The configuration does not explicitly show the implementation of IPsec ESP with IKEv2 certificates-based authentication or DTLS for the F1-C and F1-U interfaces.
    *Recommendation: Implement IPsec ESP with IKEv2 certificates-based authentication for F1-U and F1-C, and DTLS for F1-C, as mandated by 3GPP TS 33.501 [3], clause 9.1.2 (Filename: `ts_133501v180900p.pdf`).
2SCTP Association Security: The configuration specifies SCTP parameters but does not include explicit security mechanisms (e.g., IPsec, DTLS) to protect SCTP associations on N2, Xn-C, F1-C, and E1 interfaces.
    *Recommendation: Ensure that SCTP associations on N2, Xn-C, F1-C, and E1 interfaces are protected with confidentiality, integrity, and replay protection mechanisms, such as IPsec or DTLS, in accordance with 3GPP TS 33.501.
3.Enhanced Logging for Security Events: While logging is configured, the current levels ("info", "debug") might not be granular enough to capture all necessary details for security-related events and forensic analysis.
    *Recommendation: Review and enhance logging configurations to ensure that all security-relevant events are captured in accordance with O-RAN.WG11.TS.STS-R004-v11.00.pdf, clause 7.8.1.
\end{lstlisting}

\subsection{Corrected Configuration (Proposed Fix)}
\begin{lstlisting}[style=supp]
Active_gNBs = ( "gNB-Eurecom-CU");
# Asn1_verbosity, choice in: none, info, annoying
Asn1_verbosity = "none";

gNBs =
(
 {
    ////////// Identification parameters:
    gNB_ID = 0xe00;

#     cell_type =  "CELL_MACRO_GNB";

    gNB_name  =  "gNB-Eurecom-CU";

    // Tracking area code, 0x0000 and 0xfffe are reserved values
    tracking_area_code  =  1;
    plmn_list = ({ mcc = 208; mnc = 99; mnc_length = 2; snssaiList = ({ sst = 1 }) });

    nr_cellid = 12345678L;

    tr_s_preference = "f1";

    local_s_address = "127.0.0.4";
    remote_s_address = "127.0.0.3";
    local_s_portc   = 501;
    local_s_portd   = 2152;
    remote_s_portc  = 500;
    remote_s_portd  = 2152;

    # ------- SCTP definitions
    SCTP :
    {
        # Number of streams to use in input/output
        SCTP_INSTREAMS  = 2;
        SCTP_OUTSTREAMS = 2;
    };

    ////////// AMF parameters:
    amf_ip_address = ({ ipv4 = "192.168.69.131"; });

    NETWORK_INTERFACES :
    {
        GNB_IPV4_ADDRESS_FOR_NG_AMF              = "192.168.18.207";
        GNB_IPV4_ADDRESS_FOR_NGU                 = "192.168.18.207";
        GNB_PORT_FOR_S1U                         = 2152; # Spec 2152
    };
  }
);

security = {
  # preferred ciphering algorithms
  # the first one of the list that an UE supports in chosen
  # valid values: nea0, nea1, nea2, nea3
  ciphering_algorithms = ( "nea2" );

  # preferred integrity algorithms
  # the first one of the list that an UE supports in chosen
  # valid values: nia0, nia1, nia2, nia3
  integrity_algorithms = ( "nia2" );

  # setting 'drb_ciphering' to "no" disables ciphering for DRBs, no matter
  # what 'ciphering_algorithms' configures; same thing for 'drb_integrity'
  drb_ciphering = "yes";
  drb_integrity = "yes";
};

log_config :
{
  global_log_level ="info";
  hw_log_level     ="info";
  phy_log_level    ="info";
  mac_log_level    ="info";
  rlc_log_level    ="debug";
  pdcp_log_level   ="info";
  rrc_log_level    ="info";
  f1ap_log_level   ="debug";
  ngap_log_level   ="debug";
};
\end{lstlisting}

\subsection{Reflection Output (JSON)}
\begin{lstlisting}[style=supp,language=json]
{
  "OverallAssessment": "The Compliance Assessment Agent's output is correct and adheres to all specified rules, including scope control and minimal-change enforcement.",
  "Issues": [],
  "MustFixSummary": []
}
\end{lstlisting}